\documentclass[conference]{IEEEtran}
\IEEEoverridecommandlockouts
\usepackage{cite}
\usepackage{amsmath,amssymb,amsfonts}
\usepackage{algorithmic}
\usepackage{graphicx}
\usepackage{textcomp}
\usepackage{xcolor}

\usepackage{listings}

\usepackage{flushend}

\def\BibTeX{{\rm B\kern-.05em{\sc i\kern-.025em b}\kern-.08em
    T\kern-.1667em\lower.7ex\hbox{E}\kern-.125emX}}
\begin{document}

\title{The BIDS Toolbox:\\A web service to manage brain imaging datasets
\thanks{This work was supported by the Science and Technology Facilities Council (STFC), United Kingdom (grant number ST/S00209X/1).}
}

\author{
\IEEEauthorblockN{Unai Lopez-Novoa}
\IEEEauthorblockA{
\textit{Data Innovation} \\
\textit{Research Institute (DIRI)} \\
\textit{Cardiff University}\\
Cardiff, CF24 3AA, UK\\
LopezU@cardiff.ac.uk}
\and
\IEEEauthorblockN{Cyril Charron, John Evans}
\IEEEauthorblockA{
\textit{Cardiff University Brain Research} \\
\textit{Imaging Centre (CUBRIC)} \\
\textit{Cardiff University}\\
Cardiff, CF24 4HQ, UK\\
\{CharronC, EvansJ31\}@cardiff.ac.uk}
\and
\IEEEauthorblockN{Leandro Beltrachini}
\IEEEauthorblockA{
\textit{Cardiff University Brain Research} \\
\textit{Imaging Centre (CUBRIC)} \\
\textit{School of Physics and Astronomy} \\
\textit{Cardiff University}\\
Cardiff, CF24 4HQ, UK\\
BeltrachiniL@cardiff.ac.uk}
}

\maketitle

\begin{abstract}
Data sharing is a key factor for ensuring reproducibility and transparency of scientific experiments, and neuroimaging is no exception. The vast heterogeneity of data formats and imaging modalities utilised in the field makes it a very challenging problem. In this context, the Brain Imaging Data Structure (BIDS) appears as a solution for organising and describing neuroimaging datasets. Since its publication in 2015, BIDS has gained widespread attention in the field, as it provides a common way to arrange and share multimodal brain images. Although the evident benefits it presents, BIDS has not been widely adopted in the field of MRI yet and we believe that this is due to the lack of a go-to tool to create and managed BIDS datasets. Motivated by this, we present the BIDS Toolbox, a web service to manage brain imaging datasets in BIDS format. Different from other tools, the BIDS Toolbox allows the creation and modification of BIDS-compliant datasets based on MRI data. It provides both a web interface and REST endpoints for its use. In this paper we describe its design and early prototype, and provide a link to the public source code repository. 
\end{abstract}

\begin{IEEEkeywords}
Neuroscience, Neuroimaging, MRI, BIDS
\end{IEEEkeywords}

\section{Introduction}

Neuroimaging data is very heterogeneous. In its most general form, it may comprise information in plenty different formats, containing from single scalar quantities to strings and multidimensional data arrays. The wide variety of existing protocols, nomenclatures, and instruments make data sharing a demanding challenge in the field. Addressing this problem is crucial for facilitating collaborations between colleagues and centres, as well as to enhancing reproducibility and transparency of results. Moreover, it becomes a crucial organisational aspect for arranging large databases based on numerous subjects, each of them scanned with multiple imaging instruments providing complementary information of brain structure and function. Commonly found examples are the Human Connectome project \cite{VanEssen2013} in the US and the UK Biobank \cite{Bycroft2018} and WAND \cite{WAND} studies in the UK. 

To tackle this issue, Gorgolewsky et al. \cite{gorgolewski2016brain} proposed the Brain Imaging Data Structure (BIDS) format. BIDS is a community-led standard for organising and describing neuroimaging data and behavioural information, maximising their usability and, consequently, open data practices. In few years, it has found an increasingly important role in neuroimaging communities, including fMRI \cite{gorgolewski2016brain}, MEG \cite{Niso2018}, and EEG \cite{Pernet2019}. However, despite of the efforts of the community to define the standard, it has not been widely embraced by the MRI community in general. The reason, we think, is mostly based on the lack of a comprehensive and simple-to-use tool for managing and converting MRI raw data to BIDS format. Existing tools in the field (see Section~\ref{sec_dcm2bids}) lack of some key functionalities required by scientists, such as the possibility to modify an existing BIDS structure (e.g. by adding new data) or to automatically categorise the medical images without additional information other than the raw data. 

To solve this problem, we propose the BIDS Toolbox, an open source software tool that simplifies the adoption of BIDS for researchers and institutions working the field of neuroimaging. In this paper, we present the design and early prototype of a software tool for facilitating the creation and manipulation of BIDS datasets. This includes the automatic categorisation of MRI data with heuristics based on MR sequence parameters, as well as the possibility to modify existing datasets by adding new data and/or parameters. The tool can be used as a service for automated data workflows at the institution level as well as through a web interface that makes the use of the tool accessible from any modern web browser in a point-and-click manner.

The remainder of the paper is structured as follows: Section~\ref{sec_bids_software} provides a brief review of the relevant tools in the BIDS software ecosystem, highlighting the limitations that may be attempting its widespread utilisation. Section \ref{sec_toolbox} describes the BIDS Toolbox, whose performance is evaluated and presented in Section \ref{sec_eval}. Finally, Section \ref{sec_concl} draws some conclusions and describes future lines of work.



\section{BIDS software ecosystem} \label{sec_bids_software} 

\subsection{BIDS dataset creators}  \label{sec_dcm2bids}

Up to our knowledge, there are five publicly available software packages to create BIDS structures based on MRI in DICOM format, all written in Python. These tools require some external metadata in addition to the images, and use the dcm2niix~\cite{LI201647} tool for the conversion of images from DICOM to NIfTI format (as required by BIDS). The tools are: 

\begin{itemize}

\item \underline{Dcm2Bids}\footnote{Dcm2Bids - https://github.com/cbedetti/Dcm2Bids}: it allows the  conversion of one session of brain imaging for one subject at a time, with a session defined as all the acquisitions between the entry and exit of the participant in the MR scanner. It requires to set configuration options in a JSON file prior the conversion.

\item \underline{bidskit}\footnote{bidskit - https://github.com/jmtyszka/bidskit}: it permits to convert a set of several sessions for several subjects into BIDS in one go. However, it requires to arrange the DICOM files in a particular way, and to run the tool twice over the same dataset to complete the conversion, needing manual editing of a JSON configuration file between the two runs.

\item \underline{bidsify}\footnote{bidsify - https://github.com/spinoza-rec/bidsify}:  similar to bidskit, it requires to arrange the source DICOM files in a particular way prior to conversion, and filling a configuration file but in YAML format.

\item \underline{Heudiconv}\footnote{Heudiconv - https://github.com/nipy/heudiconv}: it takes DICOM files as input and produces NIfTI files arranged into structured directory layouts as output, not necessarily BIDS. It requires the user to provide a heuristic that describes the desired conversion.

\item \underline{dac2bids}\footnote{dac2bids - https://github.com/dangom/dac2bids}: similar to bidsify. It requires the manual creation of the folders structure. In addition, it only supports DICOM files from the latest Siemens scanners (VD13+). 

\end{itemize}

All these tools share the goal of creating BIDS datasets from DICOM files, but have different limitations that represent a burden for their adoption or for their integration in automated image processing pipelines. For this end, we propose the BIDS Toolbox as a software that simplifies the adoption of BIDS.

\subsection{Other tools}

In addition to the aforementioned software packages, there are other BIDS related tools  in the community. One of them is PyBIDS\footnote{PyBIDS - https://github.com/bids-standard/pybids}, a library that allows to read and extract information from a BIDS dataset using Python. Another is the BIDS Validator\footnote{BIDS Validator - https://github.com/bids-standard/bids-validator}, which is employed for checking the compliance of a given dataset with the BIDS standard and optionally with some of its extensions. This tool conforms the first sanity check in BIDS data processing workflows (e.g.~\cite{SAMPERGONZALEZ2018504}). 


\section{The BIDS toolbox} \label{sec_toolbox}

The BIDS Toolbox aims at being a software piece that is easy to integrate in existing data centres and research environments willing to adopt BIDS as a format to share neuroimaging data. To that end, we chose common design practices in software engineering and adopted a microservice design, which enables modularity and facilitates integration with other services, like an image management platform (XNAT\cite{Marcus2007}, LORI\cite{Das2012},...). 

The Toolbox functionality is exposed through a REST API and uses JSON as communication format. In the current implementation of the Toolbox, we have used and modified parts of the open source software \textit{bidskit} (described in Section \ref{sec_dcm2bids}) for some of the dataset-creation features of the toolbox, and the Flask\footnote{Flask - http://flask.pocoo.org} framework to create the web services. All the codebase is Python v3.

\subsection{REST endpoints}

The Toolbox currently exposes the following REST endpoints:

\begin{itemize}
    \item \underline{createBids}: creates a dataset with DICOM files and optional additional information about the images as input. This function creates a hidden \texttt{.bidstoolbox} file inside the dataset with Toolbox-related metadata to enable further update operations.

    \item \underline{updateBids}: updates a BIDS dataset with new DICOM files or additional information about the data. To this end, the toolbox reads the hidden \texttt{.bidstoolbox} file created by the previous function.
    
\end{itemize}

Both functions receive as input a message in JSON format with the structure defined in Listing \ref{sample-json}. The "scans" key contains an entry per scan session and subject with the path to a folder containing the DICOM files, and the "output" key is the path where the resulting BIDS dataset should be stored. We assume that these paths could be network mounted shares. 

The "metadata" key of the JSON message contains entries for additional information, e.g., "modalities" is used to describe the types of scan for the DICOM files and "datasetDescription" can be used to add a set of key/value pairs to the DatasetDescription.json file of the dataset.

\lstset{
    basicstyle=\normalfont\ttfamily,
    morestring=[b]",
    morestring=[d]'
}
\begin{lstlisting}[label=sample-json,caption=Sample JSON for The BIDS Toolbox]
{
 "scans":{
  "01":{
   "01":"/path/to/DICOMs/for/sub01/ses01",
   "02":"/path/to/DICOMs/for/sub01/ses02"
   }
 },
 "output":"/path/to/store/dataset",
 "metadata":{
  "modalities":
    [{
      "tag": "scan01",
      "modality": "anat",
      "type": "T1w"
    }],
  "datasetDescription":{
   "key01":"value01",
   "key02":"value02"
   }
  }
}
\end{lstlisting}

\begin{figure*}[htbp]
\centerline{
\includegraphics[width=0.95\textwidth]{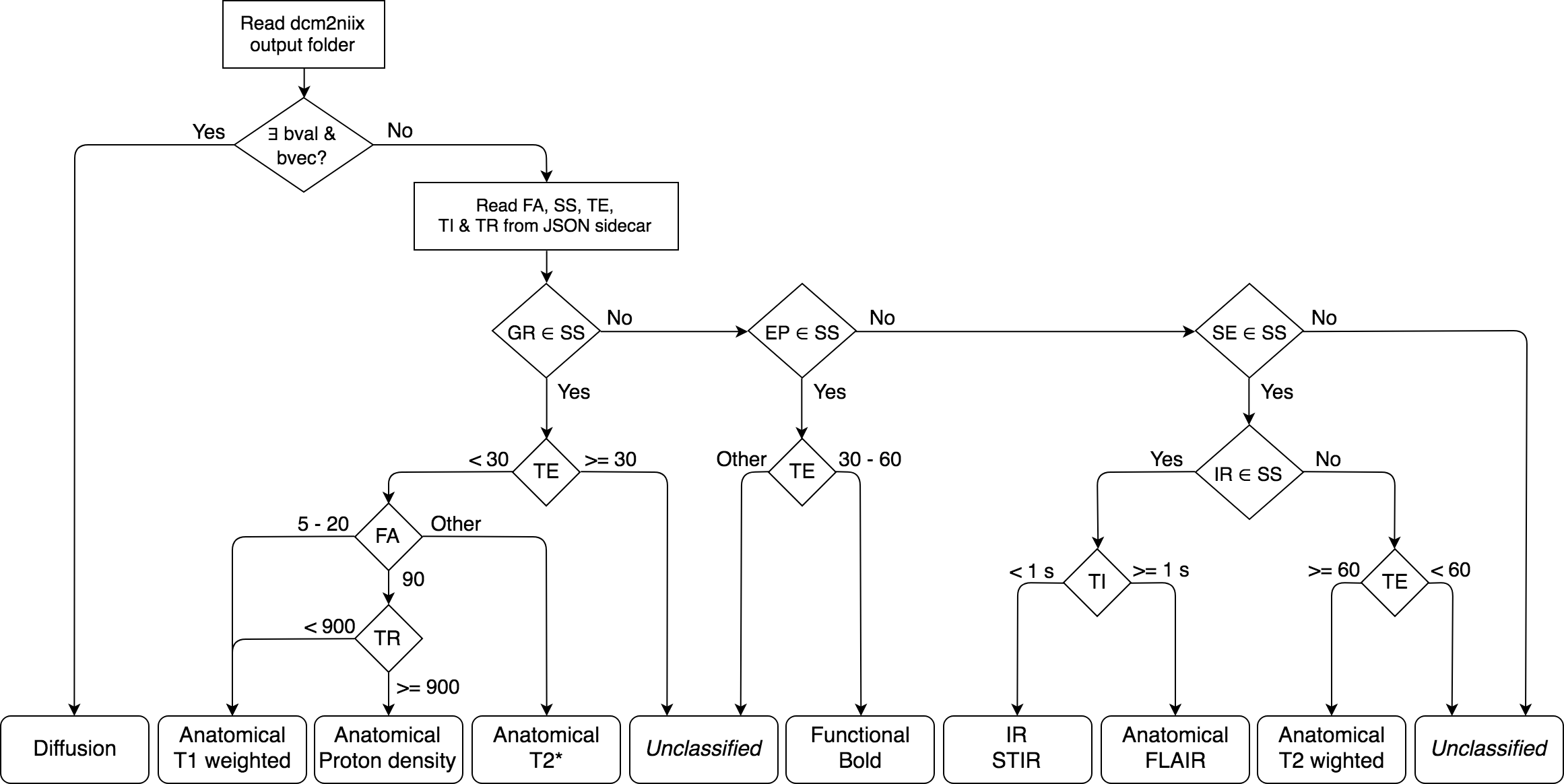}
}
\caption{Algorithm to detect scan type of a series of images. FA stands for Flip Angle, IR for Inversion Recovery, SS for Scanning Sequence, TE for Echo Time, TI for Inversion Time and TR for Repetition Time. Time values are in miliseconds.}
\label{fig_detector}
\end{figure*}

\subsection{Detecting the scan modality}

In the design of the BIDS Toolbox we assumed that the user might not know the scan modality and type for a given set of DICOM files, or that the Toolbox could be part of a processing pipeline which could not have that information. Given that this information is required to create a BIDS dataset, we developed an algorithm that infers the type of scan based on the properties of the DICOM files.

The dataset creation process in the BIDS Toolbox starts with the conversion of raw DICOM files to NIfTI using dcm2niix. After this, the Toolbox starts the scan type detection algorithm. Its logic is depicted schematically in Figure \ref{fig_detector} as a flowchart. The starting point is the output directory of dcm2niix for a particular series of scans. The Toolbox first checks in this directory if dcm2niix has created the metadata \textit{.bval} and \textit{.bvec} files with the gradient directions and diffusion weighting for the scans. If so, the modality/type of scan is defined as diffusion. If not, the Toolbox reads the Flip Angle (FA), Inversion Recovery (IR), and, if available, the Scanning Sequence (SS), Echo Time (TE), Inverstion Time (TI) and Repetition Time (TR) from the sidecar JSON file created by dcm2niix. 

Using FA, IR, SS, TE, TI and TR, the algorithm will go through a series of conditions to determine the modality and type of scans. However, it could be that the provided information is not sufficient to determine the modality and type, e.g., if the Scanning Sequence is RM (Research Mode). In these cases, the Toolbox stops the dataset creation process and returns an error message to the user with the series name that was unable to classify. The conditions and threshold values for the described algorithm have been gathered from several online sources of literature \footnote{Radiopaedia - https://radiopaedia.org/articles/mri-sequence-parameters} \footnote{MRIquestions - http://mriquestions.com/bold-pulse-sequences} and in-house expertise. 

\subsection{Web front-end}

\begin{figure*}[htbp]
\centerline{
\includegraphics[width=0.9\textwidth]{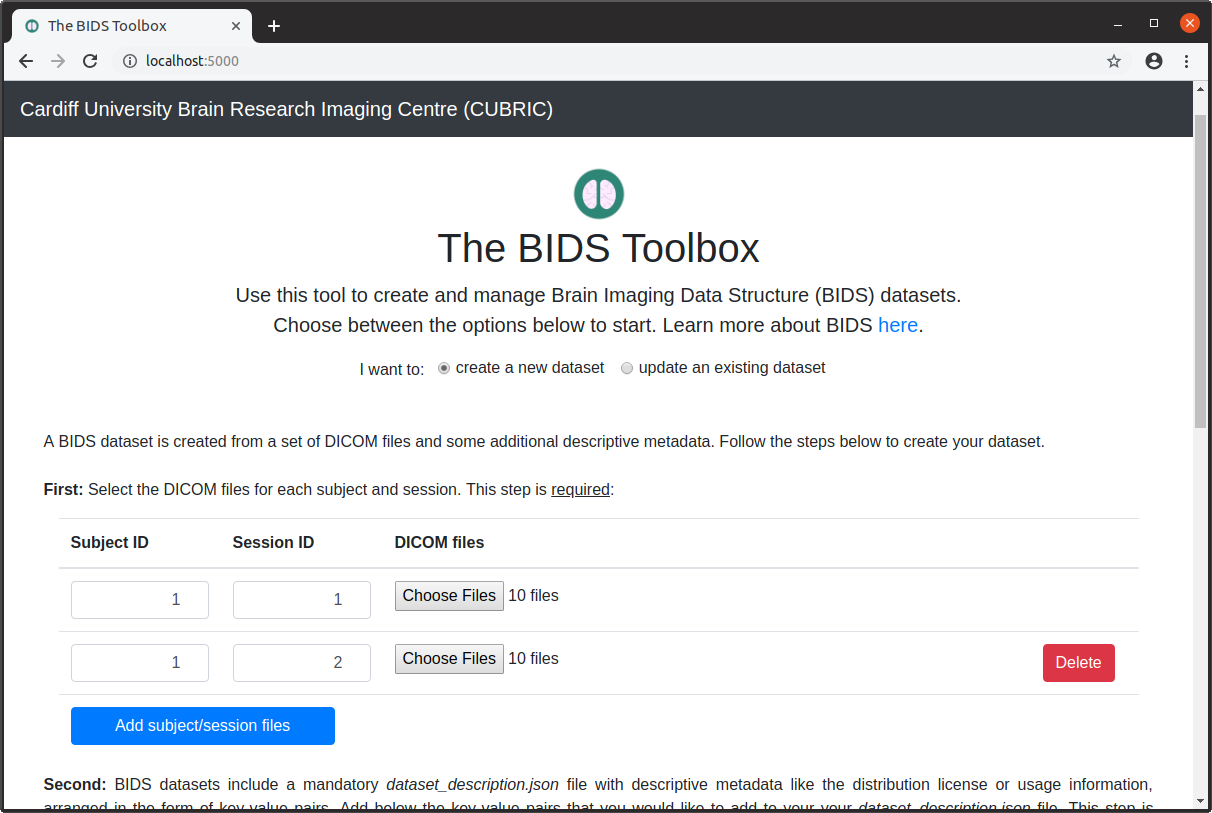}}
\caption{Screenshot of the web front-end for the BIDS Toolbox.}
\label{fig_gui}
\end{figure*}

In order to ease the use of Toolbox, we have developed a web interface that allows the users to create/update BIDS datasets using a standard web browser. It presents a simple web page that guides the user through the dataset creation/update process. We provide a screenshot of the first part of the dataset creation form in Figure \ref{fig_gui}.

The web interface works on top of the described REST API and has been developed using HTML 5, Bootstrap 4.0 and jQuery 3.2. In the current implementation of the Toolbox runs on the same Flask server as the REST services.

\section{Evaluation} \label{sec_eval}

We tested the response-time of the BIDS Toolbox in two different environments: a Virtual Machine (VM) running Ubuntu 18.10 with 1 CPU @ 2.5 Ghz and 4 GB of RAM (hosted by VirtualBox), and a workstation running Ubuntu 16.04 with an Intel Xeon CPU E5-1620 v2 (4 cores @ 3.70 GHz) and 32 GB of DDR3 RAM.

We utilised the public LGG-1p19qDeletion dataset \cite{erickson2017data}, which contains DICOM files of MRIs from pre-operative examinations performed in 159 subjects with Low Grade Gliomas (WHO grade II \& III). We used the DICOM files corresponding to the first 50 patients (code LGG-104 to code LGG-320, 727.6 MB) to test the BIDS dataset creation (\textit{createBids} function) and the DICOM files for the next 10 patients (code LGG-321 to LGG-338, 152.9 MB) to asses the inclusion of new files to the dataset (\textit{updateBids} function).

\begin{table}[htbp]
\caption{Runtime (seconds) for the BIDS Toolbox functions.}
\begin{center}
\begin{tabular}{l|c|r|c|r|}
\cline{2-5}
 & \multicolumn{2}{c|}{Virtual machine} & \multicolumn{2}{c|}{Workstation} \\ \cline{2-5} 
 & Total & \multicolumn{1}{c|}{Dcm2niix} & Total & \multicolumn{1}{c|}{Dcm2niix} \\ \hline
\multicolumn{1}{|l|}{createBids} & \multicolumn{1}{r|}{53.87} & 50.28 & \multicolumn{1}{r|}{29.11} & 27.00 \\ \hline
\multicolumn{1}{|l|}{updateBids} & \multicolumn{1}{r|}{13.18} & 10.98 & \multicolumn{1}{r|}{6.83} & 5.85 \\ \hline
\end{tabular}
\label{evaluation_tab1e}
\end{center}
\end{table}

Results for the tests are shown in Table \ref{evaluation_tab1e}. All the presented time figures are the average of 10 runs. Far from being an exhaustive performance assessment, these figures aim to show the high dependence in computation with dcm2niix: In all of the tests, more than the 80\% of the runtime is devoted to the conversion of raw DICOM files to NIfTI.  

We compiled dcm2niix with Cloudflare's implementation of zlib\footnote{Cloudflare zlib - https://github.com/cloudflare/zlib} as in all of our tests provided the best performance compared to dcm2niix's internal minigz or pgiz to create the compress NIfTI files.

\section{Conclusions \& Future work} \label{sec_concl}

We presented the BIDS Toolbox, a software tool that aims at easing the adoption of BIDS by the  neuroimaging community. It is based on the open source software \textit{bidskit} and it exposes its functionality through a REST API. The main advantages are its capability to create BIDS structures directly from DICOM data with few additional inputs, its flexibility for updating existing BIDS structures, and its easy-to-use graphical user interface. We presented an evaluation of the performance of the toolbox and described how the majority of the runtime is dominated by \textit{dcm2niix} in two different test environments. We believe that the BIDS Toolbox will facilitate to spread the use of BIDS formats within the neuroimaging community.

Future work will span in two directions: the first one is improving and validating the accuracy of the scan modality/type detection algorithm. Its accuracy will be assessed with different types of datasets and conditions, including edge cases. The second line of future work will be to improve the quality of the BIDS Toolbox as a software, moving it from a prototype status to being a production ready tool. This will imply further testing with more datasets, replacing the integrated Flask server with more stable alternatives like Gunicorn, and assessing its scalability.

The BIDS Toolbox is publicly available in CUBRIC's GitHub repository \footnote{The BIDS Toolbox - https://github.com/cardiff-brain-research-imaging-centre/bids-toolbox}.

\section*{Acknowledgment}

Authors would like to thank Greg Parker from CUBRIC for his feedback on the scan modality detection algorithm.

\bibliographystyle{IEEEtran}
\bibliography{IEEEabrv,biblio.bib}

\begin{thebibliography}{10}
\providecommand{\url}[1]{#1}
\csname url@samestyle\endcsname
\providecommand{\newblock}{\relax}
\providecommand{\bibinfo}[2]{#2}
\providecommand{\BIBentrySTDinterwordspacing}{\spaceskip=0pt\relax}
\providecommand{\BIBentryALTinterwordstretchfactor}{4}
\providecommand{\BIBentryALTinterwordspacing}{\spaceskip=\fontdimen2\font plus
\BIBentryALTinterwordstretchfactor\fontdimen3\font minus
  \fontdimen4\font\relax}
\providecommand{\BIBforeignlanguage}[2]{{%
\expandafter\ifx\csname l@#1\endcsname\relax
\typeout{** WARNING: IEEEtran.bst: No hyphenation pattern has been}%
\typeout{** loaded for the language `#1'. Using the pattern for}%
\typeout{** the default language instead.}%
\else
\language=\csname l@#1\endcsname
\fi
#2}}
\providecommand{\BIBdecl}{\relax}
\BIBdecl

\bibitem{VanEssen2013}
D.~V. Essen, S.~Smith, D.~Barch \emph{et~al.}, ``The wu-minn human connectome
  project: An overview,'' \emph{NeuroImage}, vol.~80, pp. 62--79, 2013.

\bibitem{Bycroft2018}
C.~Bycroft, C.~Freeman, D.~Petkova \emph{et~al.}, ``The uk biobank resource
  with deep phenotyping and genomic data,'' \emph{Nature}, vol. 562, pp.
  203--209, 2018.

\bibitem{WAND}
\BIBentryALTinterwordspacing
{Cardiff University Brain Research Imaging Centre (CUBRIC)}, ``Multi-scale and
  multi-modal assessment of coupling in the healthy and diseased brain,'' 2019.
  [Online]. Available:
  \url{https://www.cardiff.ac.uk/cardiff-university-brain-research-imaging-centre/research/projects/multi-scale-and-multi-modal-assessment-of-coupling-in-the-healthy-and-diseased-brain}
\BIBentrySTDinterwordspacing

\bibitem{gorgolewski2016brain}
K.~J. Gorgolewski, T.~Auer, V.~D. Calhoun, R.~C. Craddock, S.~Das, E.~P. Duff,
  G.~Flandin, S.~S. Ghosh, T.~Glatard, Y.~O. Halchenko \emph{et~al.}, ``The
  brain imaging data structure, a format for organizing and describing outputs
  of neuroimaging experiments,'' \emph{Scientific Data}, vol.~3, p. 160044,
  2016.

\bibitem{Niso2018}
G.~Niso, K.~Gorgolewski, E.~Bock \emph{et~al.}, ``Meg-bids, the brain imaging
  data structure extended to magnetoencephalography,'' \emph{Scientific Data},
  vol.~5, p. 180110, 2018.

\bibitem{Pernet2019}
\BIBentryALTinterwordspacing
C.~Pernet, S.~Appelhoff, G.~Flandin \emph{et~al.}, ``Bids-eeg: an extension to
  the brain imaging data structure (bids) specification for
  electroencephalography,'' \emph{PsyArxiv}, 2019. [Online]. Available:
  \url{https://psyarxiv.com/63a4y/}
\BIBentrySTDinterwordspacing

\bibitem{LI201647}
X.~Li, P.~S. Morgan, J.~Ashburner, J.~Smith, and C.~Rorden, ``The first step
  for neuroimaging data analysis: Dicom to nifti conversion,'' \emph{Journal of
  Neuroscience Methods}, vol. 264, pp. 47 -- 56, 2016.

\bibitem{SAMPERGONZALEZ2018504}
J.~Samper-Gonz{\'{a}}lez, N.~Burgos, S.~Bottani, S.~Fontanella, P.~Lu,
  A.~Marcoux, A.~Routier, J.~Guillon, M.~Bacci, J.~Wen, A.~Bertrand, H.~Bertin,
  M.-O. Habert, S.~Durrleman, T.~Evgeniou, and O.~Colliot, ``Reproducible
  evaluation of classification methods in alzheimer's disease: Framework and
  application to mri and pet data,'' \emph{NeuroImage}, vol. 183, pp. 504 --
  521, 2018.

\bibitem{Marcus2007}
\BIBentryALTinterwordspacing
D.~S. Marcus, T.~R. Olsen, M.~Ramaratnam, and R.~L. Buckner, ``The extensible
  neuroimaging archive toolkit: an informatics platform for managing,
  exploring, and sharing neuroimaging data,'' \emph{Neuroinformatics}, vol.~5,
  no.~1, pp. 11 -- 33, Mar 2007. [Online]. Available:
  \url{https://doi.org/10.1385/NI:5:1:11}
\BIBentrySTDinterwordspacing

\bibitem{Das2012}
\BIBentryALTinterwordspacing
S.~Das, A.~Zijdenbos, D.~Vins, J.~Harlap, and A.~Evans, ``{LORIS}: a web-based
  data management system for multi-center studies,'' \emph{Frontiers in
  Neuroinformatics}, vol.~5, p.~37, 2012. [Online]. Available:
  \url{https://www.frontiersin.org/article/10.3389/fninf.2011.00037}
\BIBentrySTDinterwordspacing

\bibitem{erickson2017data}
\BIBentryALTinterwordspacing
B.~Erickson, Z.~Akkus, J.~Sedlarand, and P.~Korfiatis, ``Data from
  lgg-1p19qdeletion. {T}he {C}ancer {I}maging {A}rchive.'' 2017. [Online].
  Available:
  \url{https://wiki.cancerimagingarchive.net/display/Public/LGG-1p19qDeletion}
\BIBentrySTDinterwordspacing

\end{thebibliography}

\end{document}